\documentclass[]{aa}
\usepackage{graphicx}
\usepackage{amsfonts}
\usepackage{amsmath}
\newcommand{\msun}{\mbox{ M}_\odot}

\begin{document}


\title{Proper motion of very low mass stars and brown dwarfs in the
  Pleiades cluster\thanks{Based on observations obtained at
    Canada-France-Hawaii Telescope}} 

\author{E. Moraux\inst{1}\and J. Bouvier\inst{1}\and J.R. Stauffer\inst{2}}

\offprints{E. Moraux}

\mail{moraux@laog.obs.ujf-grenoble.fr}

\institute{Laboratoire d'Astrophysique, Observatoire de Grenoble,
  Universit\'{e} Joseph Fourier, B.P.  53, 38041 Grenoble Cedex 9, France\\
  http://www-laog.obs.ujf-grenoble.fr/activites/starform/formation.html
  \and SIRTF Science Center, Caltech, Pasadena, CA 91125} 

\date{Received 16 November 2000/ Accepted  01 December 2000}

\authorrunning{E. Moraux et al.}
\titlerunning{Proper motion of Pleiades brown dwarfs}

\maketitle

\abstract { We report proper motion measurements for 25 very-low mass (VLM)
  star and brown dwarf (BD) candidates of the Ple\-iades cluster previously
  identified by Bouvier et al. (1998). Proper motions are measured with an
  accuracy of 9 mas/yr, compared to an expected tangential motion of about
  50 mas/yr for Pleiades members. Of the 25 candidates, 15 have a
  membership probability of 95\% or more and 7 are rejected as being field
  dwarfs. The 3 remaining candidates exhibit independent evidence for
  membership (lithium absorption or long-term proper motion). From the firm
  identification of Pleiades VLM and BD members, the cluster's substellar
  mass function is revised to $dN/dM\propto M^{-0.5}$ in the mass range
  from 0.04 to 0.3$M_\odot$.}

  
\section{Introduction}

The determination of the stellar mass function is an important challenge
for various domains of astrophysics such as, e.g., the star formation
process, the structure and evolution of the Galaxy, the dynamical evolution
of stellar clusters and stellar systems, etc... This function describes the
relative number of stars par unit mass and is usually approximated by
power-law segments of the form dN/dM $\propto$ M$^{-\alpha}$ in various
mass domains.  While its shape is relatively well constrained from
solar-type to massive stars ($\alpha\simeq 2.3-2.7$ for M$\geq$1$\msun$,
Salpeter 1955, Scalo 1998 and references therein), it is more uncertain for
low-mass stars ($\alpha\simeq$ 1.0-2.0 for 0.3$\leq$M$\leq$1$\msun$, Kroupa
2000 and references therein) and, up to a couple of years ago, was still
unexplored at very low masses in both the stellar
(0.08$\leq$M $\leq$0.3$\msun$) and substellar (M$\leq$0.08$\msun$) regimes.
The determination of the {\it substellar\/} mass function is indeed one of
the main motivations for the recent and exploding quest for free-floating
brown dwarfs (see, e.g., Oppenheimer et al. 2000 for a review).

An estimate of the mass function (MF) in the upper part of the substellar
domain ($\sim$0.04-0.08$\msun$) now exist for the Pleiades cluster (Bouvier
et al.  1998: $\alpha\simeq 0.6\pm 0.15$, Mart\'{\i}n et al.  1998, 2000:
$\alpha\simeq 0.5-1.0$, Hambly et al. 1999: $\alpha\sim 0.7$) and has
recently been derived for field brown dwarfs in the solar neighbourhood
(Reid et al.  1999: $\alpha\simeq$ 1.0-2.0 depending on the local stellar
birthrate). Determinations of the substellar mass function have also been
reported for several star forming regions (e.g.  Comeron et al. 2000,
Luhman \& Rieke 1999, Hillenbrand \& Carpenter 2000) with, however, the
additional difficulty of varying extinction within the molecular cloud
which makes more problematic the derivation of an unbiased sample of very
low-mass young objects that can be used to build the substellar MF.

Estimates of the substellar mass function of the Pleiades cluster rely on
deep, wide-field photometric surveys that identify substellar candidates
from their location in optical color-magnitude diagrams (CMD). One of the
major concern regarding these samples of very low-mass (VLM) and brown
dwarf (BD) candidates is the degree of contamination by foreground
late-type field dwarfs, which may lie in the same region of the CMD as the
Pleiades low-mass members. Bouvier et al. (1998) thus estimated at about
30\% the level of contamination of their candidate sample by field dwarfs.
Some of the BD candidates have actually been confirmed through the
``lithium test'' (Rebolo et al. 1992) but, even with the largest
telescopes, only the brightest substellar candidates are amenable to this
test (Stauffer et al.  1998). Other Pleiades BD candidates originally
identified from their location in an (I, R-I) CMD have later been rejected
on the basis of their discrepant near-IR colors (Mart\'{\i}n et al. 2000).
However, it is currently unclear whether all contaminating field stars can
be unambiguously recognized in a near-IR CMD.

A powerful way to recognize true Pleiades members among the photometric
candidates is to measure their proper motion. Proper motion studies should
allow us to pick out members with a high degree of confidence because the
cluster's peculiar motion is large compared to the non-member field stars
($\mu_{\alpha}\cos\delta = 19.15 \pm 0.23$ mas/yr, $\mu_{\delta} = -45.72
\pm 0.18$, Robichon et al. 1999) and the intrinsic velocity dispersion of
the cluster is small ($\sim 1$ mas/yr, Van Leeuwen 1980). Such a large
motion can be measured from sets of sharp images separated by only a few
years and as soon as a candidate is recognized as a Pleiades member on the
basis of its kinematics, its status (VLM star or BD) directly follows from
its photometric properties.

We therefore obtained in September 1999 new images for 25 of the 26 VLM
star and BD candidates of the Pleiades originally observed by Bouvier et
al.  (1998) in December 1996. In Section 2, we describe how proper motion
was derived for each candidate from the 2 sets of images separated by
nearly 3 years. In most cases, the results presented in Section 3 allow us
to unambiguously identified Pleiades members and they are compared with
other diagnostics of Pleiades membership. In Section 4, we discuss whether
previous estimates of the Pleiades substellar function has to be revised in
the light of these new results and briefly discuss the kinematics of very
low-mass members of the cluster with respect to their formation mechanism.

\section{Observations and astrometric reduction}

The first set of images, from which VLM and BD candidates were identified,
was obtained by Bouvier et al. (1998) in December 1996 with the CFHT 8K
wide-field camera. Of the 26 identified candidates, 25 were reobserved in
September 1999 with the CFHT 12K camera. Both instruments have the same
pixel size (0.21\arcsec) and the images were obtained under comparable
seeing conditions (0.9\arcsec and 0.7\arcsec, respectively). In 1999, the
exposure time was adjusted so as to provide similar signal-to-noise ratio
on the candidates as on the 1996 images. Astrometry was performed on I-band
images from the 2 epochs.

The principle of the astrometric procedure we used to measure the proper
motion of the candidates is as follows. We first measure the position
(x$_{96}$,y$_{96}$), (x$_{99}$,y$_{99}$) of the candidate on the two sets
of images using IRAF/CENTER. On the same images, we also measure the
positions of typically 10 point-like objects, presumably background field
stars, located within an angular radius of about 3\arcmin\ from the
candidate on the CCD. These stars are then used as relative astrometric
references to compute the spatial transformation function that maps 1999
coordinates to the 1996 reference frame using the IRAF/IM\-MATCH package. The
rms uncertainty associated to the spatial transformation, $\sigma_t$, is
computed by the IRAF task as well as the residuals for each astrometric
reference star. In case the residuals for a reference star is significantly
larger than $\sigma_t$, the object is discarded and the transformation
recomputed.

This transformation is applied to (x$_{99}$,y$_{99}$) in order to project
the 1999 coordinates of the candidate in the 1996 reference frame:
$(x_{99},y_{99}) \longrightarrow (x_{99 \rightarrow 96},y_{99 \rightarrow
  96})$. The displacement of the object in pixels between 1996 and 1999 is
simply given by:

$$\Delta x= x_{99\rightarrow 96}-x_{96}=(\Delta t)\mu_{x}$$
$$\Delta y= y_{99\rightarrow 96}-y_{96}=(\Delta t)\mu_{y}$$

where $\Delta t$ is the time lag between the two epochs and $\mu_{x}$,
$\mu_{y}$ are the relative proper motions along the x and y axes of the CCD
in the 1996 reference frame. The precision of the measurement is
$\sigma_{\Delta x}^2 = \sigma_{\Delta y}^2 = 2\sigma_{x,y}^2 +
\sigma_{t}^2$, where $\sigma_{x,y}$ is the rms error on the measurement of
the position of the candidate on the CCD and $\sigma_t$ the error
associated with the spatial transformation from one epoch to the other.  We
evaluated $\sigma_{x,y}\simeq 0.07$ pixels by measuring the positions of
the same objects on consecutive I-band images of the same field. Tests were
run with the IRAF/GEOMAP package in order to minimize $\sigma_t$. We found
that a third degree polynomial transformation yields the best results with
$\sigma_t \simeq$ 0.05 pixels. We verified a posteriori that the proper
motion measured for each candidate was not unduely sensitive either to the
degree of the polynomial transformation or to the number of astrometric
reference stars used to compute the transformation.

The displacement $(\mu_{x}, \mu_{y})$ in pixels/yr has still to be
converted into a displacement on the sky $(\mu_{\alpha}\cos\delta,
\mu_{\delta})$ in mas/yr by calibrating the spatial scale and orientation
of the 1996 coordinates system. The calibration was done using PRIAM
(Proc\'edure de R\'eduction d'Images AstroM\'etriques), a software package
developed by A.Fienga and J.Berthier at the IMCCE (Institut de M\'ecanique
C\'eleste et de Calcul des \'Eph\'em\'erides). This software calculates the
transformation which converts pixel positions $(x,y)$ into J2000 celestial
coordinates $(\alpha,\delta)$ for a set of astrometric standards. For each
candidate field, we measured on the image the position of typically 10
stars belonging to the USNO2 catalogue whose astrometric accuracy is of
order 500 mas.  PRIAM then computes the coefficients A0,..,B2 of the
transformation:

$$\alpha = A0+A1\,\,x+A2 \,\, y$$
$$\delta = B0+B1\,\,x+B2 \,\, y$$

from which we obtain: 

$$\mu_{\alpha}\cos\delta=A1\,\,\mu_{x}+A2\,\,\mu_{y}$$
$$\mu_{\delta}=B1\,\,\mu_{x}+B2\,\,\mu_{y}$$

The rms error on these coefficients is $\sim 0.1$ mas/pixel, and can be
neglected compared to $\sigma_{x,y}$ and $\sigma_{t}$. Hence, the final
uncertainty on the proper motion measurement is

$$\sigma_{\mu_{\alpha}\cos\delta}^2 \sim \sigma_{\mu_{\delta}}^2 \sim
\frac{pix^2}{\Delta t^2} \,(2\sigma_{x,y}^2+\sigma_{t}^2)$$

which typically amounts to 9 mas/yr rms for a pixel scale $pix = 205$ mas
and $\Delta t$ = 2.8 yr. 

\section{Results}

\begin{figure}[htbp]
  \begin{center}
    \leavevmode
    \includegraphics[width=0.9\hsize]{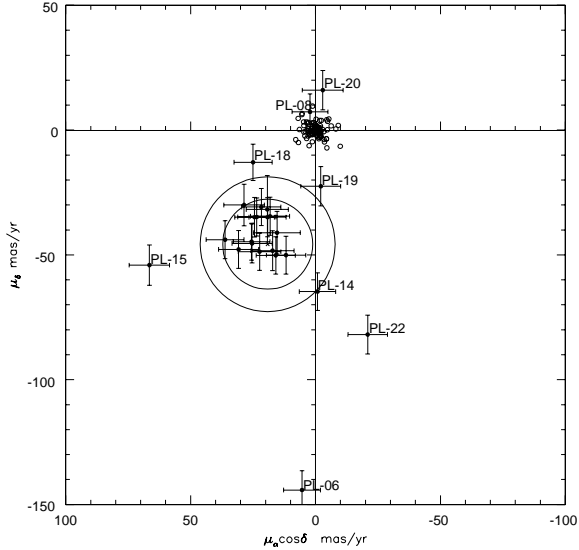}
  \end{center}
    \caption{ 
      Proper motion vector point diagram for objects located in Pleiades
      fields. Filled dots show VLM and BD candidates and open circles the
      field stars used as astrometric references. North is up and East is
      left. The cluster proper motion is indicated by a cross and two
      circles centered on this value are drawn. The radius is equal to
      $2\sigma$ and $3\sigma$ for the small and the big circle respectively
      where $\sigma$ corresponds to the 2D-gaussian dispersion of the
      candidates in this diagram (see text).}
    \label{ppmfig}
\end{figure}

The proper motions of the 25 Pleiades VLM star and BD candidates are listed
in Table~\ref{ppmtable} and illustrated in a vector point diagram (VPD) in
Figure~\ref{ppmfig}. The field stars used as astrometric references scatter
around the origin of the diagram since they have been selected on the basis
of a negligible proper motion. In contrast, most of the Pleiades
candidates fall within 3$\sigma$ of the expected cluster's mean motion.
Following the method outlined by Sanders (1971) we compute a membership
probability for each candidate as follows:
$$
p=\frac{f_{c}(\mu_{\alpha}\cos\delta,\mu_{\delta})}{f_{f}(\mu_{\alpha}\cos\delta,\mu_{\delta})+f_{c}(\mu_{\alpha}\cos\delta,\mu_{\delta})}$$
where $f_{c}(\mu_{\alpha}\cos\delta,\mu_{\delta})$ and
$f_{f}(\mu_{\alpha}\cos\delta,\mu_{\delta})$ are the vector point
distributions of the members and field stars respectively. We
assumed $f_{c}$ is a bivariate normal function and we found
\begin{equation*}
\begin{split}
f_{c}(\mu_{\alpha}\cos\delta,\mu_{\delta})&= 0.04  
\exp\Big\{-\frac{1}{2}\Big[\Big(\frac{\mu_{\alpha}\cos\delta-21.5}{7.7}\Big)^2\\
&+\Big(\frac{\mu_{\delta}+40.1}{8.5}\Big)^2\Big]\Big\}
\end{split}
\end{equation*}
by fitting a 2-D gaussian, plus an assumed uniform distribution for field
stars, on the vector-points distribution of the VLM and BD candidates. The
field stars distribution is then $f_{f}=3.3 10^{-4}$. We also tried to
reproduce the distribution of both Pleiades candidates and astrometric
references in the VPD by adding a 2-D gaussian centered on (0,0) to
the fitting function. This leads to a dispersion of $\sim 4.5$ mas/yr for
the astrometric references. Membership probabilities computed for the
candidates by either method are very similar because the distributions of
candidates and astrometric references do not overlap in the VPD.

\begin{table*}[htbp]
    \centering
    \leavevmode
    \caption{Proper motion of Pleiades VLM and BD candidates. The Pleiades cluster proper motion is $\overline{\mu_{\alpha}\cos\delta}=19.15\pm0.23$ mas/yr and $\overline{\mu_{\delta}}=-45.72\pm0.18$ mas/yr.}
    \begin{tabular}{llcccc} \hline 
Name& Other id.&$\mu_{\alpha}cos\delta \pm \sigma_{\mu_{\alpha}cos\delta}$ &$\mu_{\delta} \pm \sigma_{\mu_{\delta}}$ &&\\
\rule[-2.5ex]{0cm}{5ex} &&$(mas/yr)$&$(mas/yr)$&\raisebox{0.75ex}[0pt]{\hbox{$\displaystyle{(\mu_{\alpha}cos\delta - \overline{\mu_{\alpha}cos\delta})}\over{\displaystyle\sigma_{\mu_{\alpha}cos\delta}}$}}&\raisebox{0.75ex}[0pt]{\hbox{$\displaystyle{(\mu_{\delta} - \overline{\mu_{\delta}})}\over{\displaystyle\sigma_{\mu_{\delta}}}$}}\\
\hline
CFHT-PL-1&& $22.4\pm8.5$ & $-48.7\pm7.5$& 0.4& 0.4\\
CFHT-PL-2 && $15.4\pm9.3$ & $-41.1\pm8.5$&0.4& 0.5\\
CFHT-PL-3$^\dagger$  & HHJ22 & -- &  -- & & \\
CFHT-PL-4 && $17.2\pm8.6$ & $-48.3\pm8.0$ &0.2& 0.3\\
CFHT-PL-5 && $18.2\pm7.8$ & $-34.5\pm7.6$&0.1&1.5\\
CFHT-PL-6&& $5.4\pm7.4$ & $-144.2\pm7.7$&1.9&12.8\\
CFHT-PL-7 && $19.2\pm7.4$ & $-34.9\pm7.8$ &0.01&1.4\\
CFHT-PL-8 && $2.21\pm7.2$ & $7.31\pm7.2$&2.35&7.4\\
CFHT-PL-9$^\ddagger$ && $19.3\pm8.4$ & $-31.8\pm13.6$&0.02&1.0\\
CFHT-PL-10 && $28.6\pm8.1$ & $-30.0\pm8.3$&1.2&1.9\\
CFHT-PL-11 & Roque 16 & $25.4\pm8.0$ & $-45.4\pm7.8$&0.8&0.04\\
CFHT-PL-12 && $36.2\pm7.6$ & $-43.9\pm7.6$&2.2&0.2\\
CFHT-PL-13 & Teide 2 & $11.8\pm7.9$ & $-50.1\pm7.6$&0.9&0.6\\
CFHT-PL-14 && $-0.8\pm7.3$ & $-64.7\pm7.5$&2.7&2.5\\
CFHT-PL-15 && $66.5\pm8.1$ & $-54.1\pm8.1$&5.8&1.0\\
CFHT-PL-16 && $21.7\pm7.8$ & $-30.8\pm7.4$&0.3&2.0\\
CFHT-PL-17 && $30.8\pm8.0$ & $-47.8\pm7.6$&1.5&0.3\\
CFHT-PL-18 && $25.0\pm7.6$ & $-12.9\pm7.3$&0.8&4.5\\
CFHT-PL-19 && $-2.1\pm8.0$ & $-22.5\pm7.9$&2.6&2.9\\
CFHT-PL-20 && $-2.9\pm8.2$ & $16.0\pm7.9$&2.7&7.8\\
CFHT-PL-21 & Calar 3 & $23.5\pm7.6$ & $-34.8\pm7.6$&0.6&1.4\\
CFHT-PL-22 && $-20.9\pm7.9$ & $-81.9\pm7.8$&5.1&4.6\\
CFHT-PL-23 && $15.9\pm7.8$ & $-50.2\pm7.5$&0.4&0.6\\
CFHT-PL-24 & Roque 7 &$24.4\pm7.9$ & $-34.9\pm7.9$&0.7&1.4\\
CFHT-PL-25 && $25.6\pm7.3$ & $-44.7\pm7.4$&0.9&0.1\\
\hline 
\end{tabular}

$^\dagger$: CFHT-PL-3 is located close to the edge of a CCD and at the edge
of the mosaic's field of view. This leads to large distorsions that prevent
us from deriving a reliable measurement of its proper motion. \hfill\hfill

$^\ddagger$: The measurement error on $\mu_{\delta}$ is larger
than average for CFHT-PL-9 due to charge transfer problems on one of the
CCDs of the mosaic that smears the stellar profile.\hfill\hfill

    \label{ppmtable}
\end{table*}


The membership probability of the candidates is listed in
Table~\ref{membership}. We also list in this table other diagnostics of
Pleiades membership obtained by Stauffer et al. (1998) and Mart\'{\i}n et
al. (2000), namely: EW(Li), EW(H$_\alpha$), radial velocity, and (I-K)
index. We consider the 15 candidates for which we found $p\geq$ 95\% as
Pleiades members. It is interesting to note that two of these confirmed
members (CFHT-PL-7 a very low-mass Pleiades star, and CFHT-PL-25, the least
massive brown dwarf) do not exhibit H$\alpha$ in emission according to
Martin et al. (2000). In addition to these 15 highly probable Pleiades
members, CFHT-PL-12 has a membership probability of 88\% and lithium has
been detected in its spectrum (Stauffer et al. 1998) which makes it a bona
fide Pleiades brown dwarf. We could not measure the proper motion of
CFHT-PL-3 (= HHJ 22) because it is located both at the edge of a CCD and at
the edge of the camera field of view which makes the computation of the
spatial transformation between the two epochs unreliable. However, the
long-term proper motion of HHJ 22 has been measured by Hambly et al. (1993)
with an epoch difference of nearly 40 years and indicates highly probable
Pleiades membership.  We therefore consider this candidate as a cluster
member.

The remaining 8 candidates (CFHT-PL 6, 8, 14, 15, 18, 19, 20, 22) have very
low membership probabilites ($p\leq$20\%). Of these, CFHT-PL-14 and 18 had
already been rejected as a non-members based on the absence of lithium
absorption in their spectrum and CFHT-PL-19, 20, and 22 were suspected
non-members based on their peculiar location in near-IR CMDs. The low
membership probability we measure for these objects confirmed that they are
non-members. The 3 remaining candidates (CFHT-PL 6, 8, and 15) deserve
further discussion.

CFHT-PL-6's proper motion is high and more than 3$\sigma$ away from the
cluster's mean motion. On the one hand, CFHT-PL-6 does not exhibit
H$\alpha$ emission (Mart\'{\i}n et al.  2000) but this alone does not rule
out membership since its twin, CFHT-PL-7, also lacks H$\alpha$ emission but
is a confirmed Pleiades member from its proper motion.  On the other hand,
CFHT-PL-6 lies significantly above the Pleiades ZAMS which led Bouvier et
al. (1998) to suggest that it could be a nearly equal mass binary which, in
turn might affect its short term proper motion. If it is indeed an equal
mass binary, then the components of CFHT-PL-6 would be substellar and
therefore amenable to the lithium test. We also find that CFHT-PL-6 was
about 0.1 magnitude brighter in the I-band in 1996 than in 1999. This level
of photometric variablity is typical of late-M dwarfs (Mart\'{\i}n et al.
1996) but might also occur in substellar objects (Bailer-Jones \& Mundt
1999). Based primarily on its highly discrepant proper motion and pending
additional observations of this object, we tentatively conclude that
CFHT-PL-6 is most likely a foreground M dwarf and not a cluster member.

CFHT-PL-8 has a vanishingly small membership probability with a proper
motion close to that of background field stars. It had previously been
considered as a probable member based on EW(H$\alpha$) = 14.6\AA, a
spectral type of dM5.6 and IR colors consistent with membership
(Mart\'{\i}n et al.  2000).  These properties, however, are not
inconsistent with it being a low-mass field dwarf. Even the extremely short
rotational period of 0.401 days recently derived by Terndrup et al. (1999)
is not unexpected for very low-mass dwarfs (Delfosse et al. 1998). We thus
conclude that this candidate is not a cluster member.

CFHT-PL-15 is puzzling in several ways. There is little doubt that this
late-type object (Sp.T. M7) is a Pleiades brown dwarf since lithium has
been detected in its spectrum (Stauffer et al. 1998).  Yet, the tangential
motion we measure is clearly different from that of Pleiades members.
Moreover, Stauffer et al. (1998) find that it exhibits unusual colors that
locate it slightly below the Pleiades ZAMS.  From the analysis of HST
images, Mart\'{\i}n et al. (2000) found evidence for high residuals after
PSF subtraction, which might indicate the presence of an unresolved
companion at a separation less than 0.22\arcsec and about 3 magnitude
fainter than CFHT-PL-15. It is thus conceivable that the short-term
peculiar tangential motion we measure results from orbital motion in a
binary system or photometric variability of one or both components. The
latter appears more likely since, with an estimated substellar mass, the
maximum displacement of the photocenter due to orbital motion would be less
than 25 mas over 3 years while we measure $\sim 135$ mas. However,
CFHT-PL-15 does not seem to be a photometric binary from its location in a
color-magnitude diagram. For the time being, we thus consider CFHT-PL-15 as
a highly probable Pleiades brown dwarf based on the presence of lithium in
its spectrum.

\begin{table*}[htbp]
    \centering
    \small
    \caption{Candidates membership criteria. The $EW(Li)$ are from Stauffer
    et al. (1998) and the others tests have been performed by Mart\'{\i}n et
    al. (2000). In column 8, we give the membership probability of the
    candidates based on their proper motion. Our final assessment regarding
    membership is indicated column 9.}
    \begin{tabular}{llcccc|cccc}
\hline
Name&other id.&$EW(Li)$&$H_{\alpha}$&$V_{rad}$&I-K&\multicolumn{3}{c}{Membership}\\ 
&&(\AA)&&&&Martin et al.& prob. (\%)&Conclusion\\
\hline
CFHT-PL-1 &&&yes&&yes&yes&98.6&yes\\
CFHT-PL-2 &&&yes&&yes&yes &98.9&yes\\
CFHT-PL-3 &HHJ22 &&&&yes&yes& -- &yes\\
CFHT-PL-4 &&&&&yes&yes? &98.6&yes\\
CFHT-PL-5 &&&yes&&yes&yes &98.8&yes\\
CFHT-PL-6 & &&no&&yes&yes?&$\ll$1&no?\\
CFHT-PL-7 &&&no&&yes&yes?&98.9&yes\\
CFHT-PL-8 &&&yes&&yes&yes&$\ll$1&no\\
CFHT-PL-9 && $<0.05$ &yes&yes&yes&yes&98.4&yes\\
CFHT-PL-10 && $<0.05$ &yes&yes&yes&yes&95.7&yes\\
CFHT-PL-11 &Roque 16 & 0.5 &yes&yes&yes&yes&98.7&yes\\ 
CFHT-PL-12 && 0.8 &yes&yes&yes&yes&87.7&yes\\
CFHT-PL-13 &Teide 2 & 0.6 &yes&yes&yes&yes &96.7&yes\\
CFHT-PL-14 && $<0.1$ &no&no&&no &1.3&no\\ 
CFHT-PL-15 && 0.5 &yes&yes&yes&yes&$\ll$1&yes\\ 
CFHT-PL-16 &&1.2&yes&yes&yes&yes &98.1&yes\\
CFHT-PL-17 &&&yes&&yes&yes &96.1&yes\\
CFHT-PL-18 &&no&yes&&yes&no &17.6&no\\
CFHT-PL-19 &&&&&no&no?&4.0&no\\
CFHT-PL-20 &&&no&&no&no &$\ll$1&no\\
CFHT-PL-21 &Calar 3 &yes&yes&yes&yes&yes&98.7&yes\\ 
CFHT-PL-22 &&&&&no&no &$\ll$1&no\\
CFHT-PL-23 &&&&&yes&yes &98.1&yes\\
CFHT-PL-24 &Roque 7 &&&&&yes&98.7&yes\\
CFHT-PL-25 &&&no&&yes&yes&98.7&yes\\
\hline
\end{tabular}


    \label{membership}
\end{table*}

\section{Discussion}

The firm identification of Pleiades VLM stars and brown dwarfs from their
kinematics provides a clean, albeit small, sample of substellar objects
from which a more reliable estimate of the substellar mass function of the
cluster can be derived. In addition, these are the first proper motion
measurements obtained with an accuracy of better than 10 mas/yr for very
low mass Pleiades members and this allows us to start to investigate the
intrinsic velocity distribution of brown dwarfs in the cluster. These two
aspects are briefly discussed below after the status of the objects
contaminating the photometric sample is investigated.

\subsection{ The nature of the contaminating objects}

Seven objects of the original photometric sample are found to be probable
non members. These objects are located in the same region of the (R-I, I)
CMD as bona fide Pleiades VLM stars and brown dwarfs. Their proper motions
in Figure~\ref{ppmfig} do not seem to be uniformly distributed but tend to
scatter over the lower left quadrant of the vector-point diagram.  In order
to investigate the status of the contaminating objects we used the
kinematical model of the Galaxy developed by Robin \& Cr\'ez\'e (1986).
From the model, we constructed a synthetic sample of field stars assumed to
be observed in the galactic direction of the Pleiades cluster and covering
a magnitude range $I=16-20$ and a color range $R-I=1.7-2.7$. The sample
thus computed contains nearly 10$^4$ stars of which 13 lie in the same
location of the CMD as the Pleiades VLM and BD candidates. According to the
model, these 13 objects are late M-dwarfs (M6-M9) distributed over a
distance between 70 and 170 pc and their proper motions indeed scatter
preferentially over the lower left quadrant of the VPD with an amplitude up
to 120 mas/yr.  The same eccentric distribution of proper motions is seen
for the whole sample of 10$^4$ stars and results from the effect of
galactic differential rotation in the direction of the Pleiades cluster.
We thus conclude that the photometric candidates rejected as being non
Pleiades member based on their proper motion are a mixture of foreground
and background late-M dwarfs of the galactic disk.

\subsection{ The Pleiades substellar IMF}

From accurate proper motion measurements, we have identified 7 out of 25
VLM and BD candidate members as probable field stars. In addition, the
lowest mass candidate of Bouvier et al.'s (1998) sample, CFHT-PL-26, was
not included in this study but Mart\'{\i}n et al. (2000) classified it as a
non-member based on its spectral properties (discrepant pseudo-continuum
indices and lack of H$_\alpha$ emission). The overall level of
contamination of the photometric sample by field dwarfs is thus $8/26 =
31\%$, close to the original estimate of Bouvier et al.  (1998) based on
statistical arguments.

Do these results modify the earlier estimate of the substellar mass
function of the Pleiades cluster? Of the 18 photometric candidates with a
mass between 0.04 and 0.08$M_\odot$ in the original sample (CFHT-PL-9 to
26), 6 are rejected here as non-members. Thus counting 12 confirmed objects
in this mass range, a power law fit to the mass function between
0.3$M_\odot$ and 0.04$M_\odot$ yields: $dN/dM \propto M^{-0.51 \pm 0.15}$,
i.e., slightly shallower than the earlier estimate ($dN/dM \propto M^{-0.60
  \pm 0.15}$). This slight revision is not very significant considering
that the errors are still dominated by Poisson noise from small samples.

These results do not account for binarity. CFHT-PL-6 (whose membership is
however in doubt, see above), 12, and 16 appear to lie on the binary
sequence of the cluster in a color-magnitude diagram. If these 3 objects
are nearly equal mass cluster binaries, one finds $dN/dM\propto M^{-0.65
  \pm 0.15}$.

\subsection{Kinematics of Pleiades VLM stars and BDs}

\begin{figure}[htbp]
  \begin{center}
    \leavevmode
    \parbox{0.5\hsize}{\includegraphics[width=\hsize]{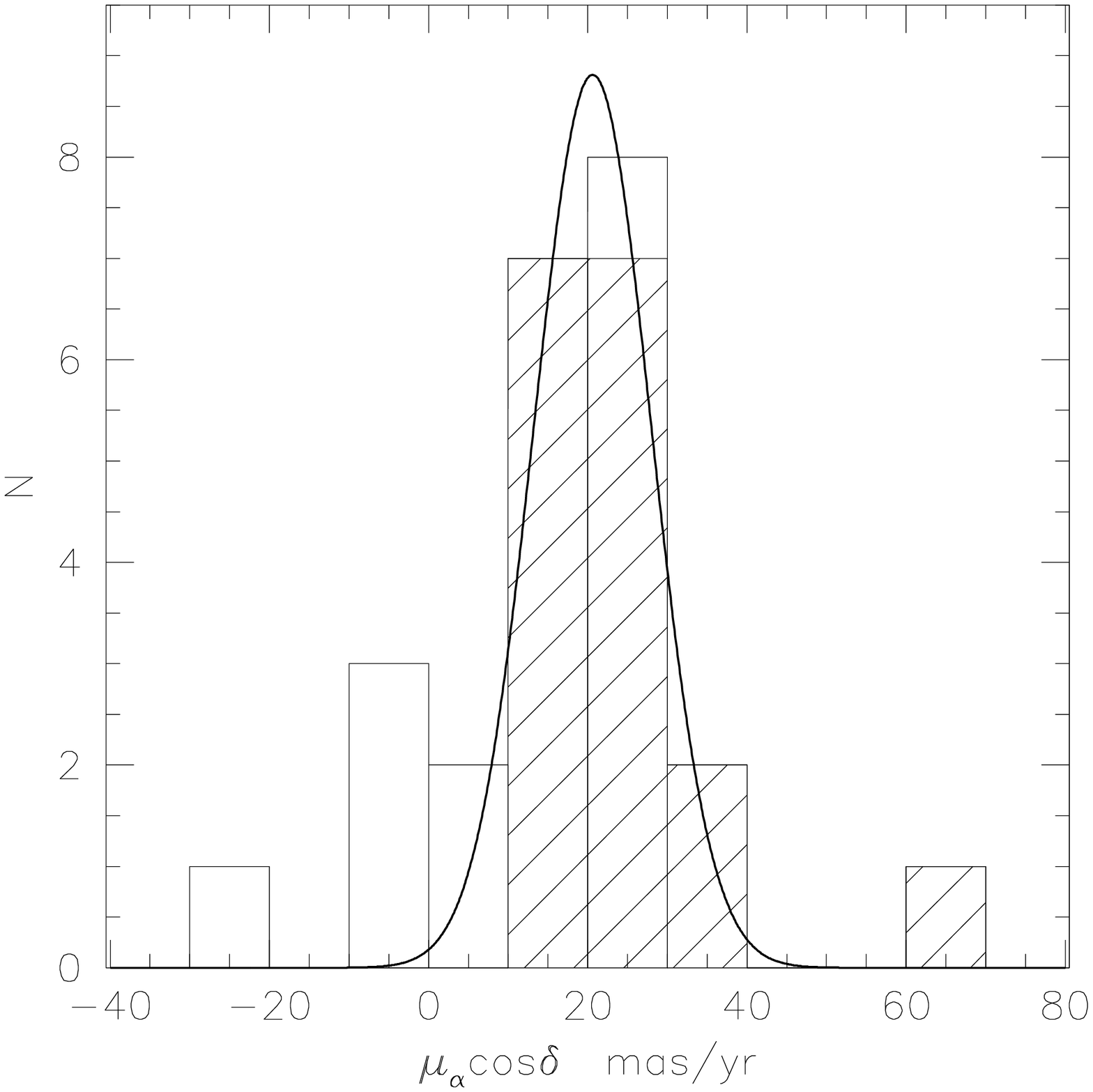}}\hfill\parbox{0.5\hsize}{\includegraphics[width=\hsize]{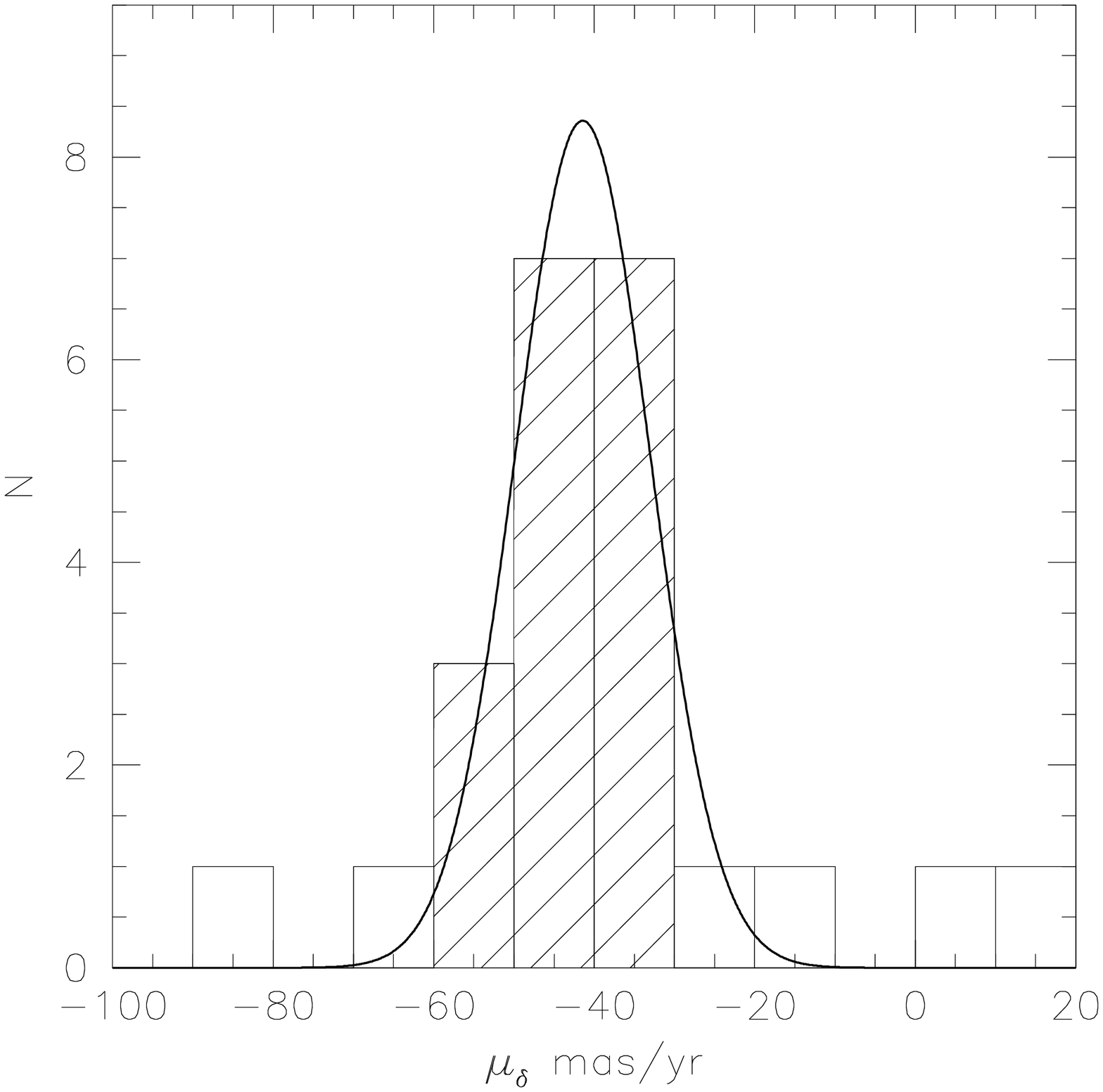}}
  \end{center}
    \caption{
      Proper motion distributions for program stars. The hatched histogram
      refers to candidates confirmed as being Pleiades members and the
      solid curve is a gaussian fit to their proper motion distribution.}
    \label{ppmhisto}
\end{figure}

The $\mu_{\alpha}\cos \delta$ and $\mu_{\delta}$ distributions of the
Pleiades VLM and BD candidates are illustrated in Fig. \ref{ppmhisto}. A
gaussian fit to the distributions of the confirmed Pleiades members leads
to dispersions of $\sigma_{\mu_{\alpha}\cos \delta} = 7.2$ mas/yr and
$\sigma_{\mu_{\delta}} = 8.5$ mas/yr that are similar to the expected
measurement error. Hence, we find no evidence for an intrinsic dispersion
in the distribution of tangential velocities of the VLM and BD Pleiades
members. This is consistent with the hypothesis that these VLM and
substellar objects formed in the same way as stellar cluster members did,
i.e., from the collapse of isolated, very low mass molecular cloud cores.
Indeed, the internal dispersion of Pleiades stars is $\sim 1$ mas/yr (Van
Leeuwen, 1980).

However, owing to the difficulty of triggering the collapse of molecular
cores whose mass is much lower than the typical Jeans mass ($\simeq 0.7
M_\odot$, Clarke et al.  2000), alternative models have been proposed for
the formation of brown dwarfs. Burkert et al. (1997) have shown that
collapsing molecular cores are prone to multiple fragmentation that
eventually leads to the formation of small-N protoclusters, including a
number of very low mass fragments. Following dynamical interactions within
the protocluster, the least massive fragments are preferentially ejected
with typical velocities of order of a few km/s (Burkert, priv.  comm.) and
may thus become isolated brown dwarfs (Klessen \& Burkert 2000).
Alternatively, Lin et al. (1998) proposed that isolated brown dwarfs may
form as a result of an encounter between protostars with massive disks. The
encounter leads to the formation of an unbound tidal tail which contains
part of the initial disk's mass and may later condensed to form an isolated
substellar object. Here again, the typical ejection velocities are of order
of a few km/s.

At a distance of 118 pc, a tangential velocity of 1 km/s corresponds to a
tangential motion of about 2 mas/yr. Proto brown bwarfs ejected with this
velocity in the early stages of cluster formation, about 120 Myr ago, would
have now drifted by several tens of degrees away from the cluster. If these
BD formation models are correct, the sample we have observed close to the
cluster's center must represent a tiny fraction of the primordial BDs,
namely those which populated the low tail of the distribution of ejection
velocities. In such a case, the slope of the substellar mass function of
the Pleiades cluster might currently be largely underestimated.

\section{Conclusion}

A sample of very low-mass stars and brown dwarfs have been firmly
identified in the Pleiades cluster on the basis of their proper motion.
With 12 confirmed objects with a mass less than 0.08$M_\odot$ distributed
over an area of 2.5 square degrees close to the cluster's center, the
Pleiades mass function is estimated to be $dN/dM \propto M^{0.51 \pm 0.15}$
in the mass domain ranging from 0.04 to 0.3$M_\odot$. Taking individual
components of suspected substellar binaries into account leads to $dN/dM
\propto M^{0.65 \pm 0.15}$. The main source of uncertainty on these
estimates now lies in the unknown radial distribution of brown dwarfs
relative to stars in the cluster. As a group, the identified brown dwarfs
exhibit no intrinsic velocity dispersion. This suggests that they have
formed from the collapse of isolated low mass molecular cores. However, we
cannot rule out that this sample of brown dwarfs represent only a tiny
fraction of the brown dwarf population originally formed in the cluster and
which might have escaped since then either due to dynamical ejection at the
time of their formation or during the subsequent dynamical evolution of the
cluster.  Additional studies covering a much larger fraction of the
cluster's area are needed to settle this issue.

\begin{acknowledgements}
We thank A. Fienga for allowing us access to PRIAM before public release, 
A. Robin for help in using the Besancon's galactic model, and C. Clarke for
discussions on cluster kinematics. 
\end{acknowledgements}

\end{document}